\documentclass[12pt]{article}
\begin{document}
\begin{center}
{\large\bf NON-SINGULAR COSMOLOGY IN MODIFIED GRAVITY} \vskip 0.3
true in {\large J. W. Moffat} \vskip 0.3 true in {\it The
Perimeter Institute for Theoretical Physics, Waterloo, Ontario,
N2L 2Y5, Canada} \vskip 0.3 true in and \vskip 0.3 true in {\it
Department of Physics, University of Waterloo, Waterloo, Ontario
N2L 3G1, Canada}
\end{center}
\begin{abstract}%

A non-singular cosmology is derived in modified gravity (MOG) with
a varying gravitational coupling strength $G(t)=G_N\xi(t)$.
Assuming that the curvature $k$, the cosmological constant
$\Lambda$ and $\rho$ vanish at $t=0$, we obtain a non-singular
universe with a negative pressure, $p_G < 0$.
The universe expands for
$t\rightarrow \infty$ according to the standard radiation and
matter dominated solutions. The thermodynamical arrow of time reverses at $t=0$
always pointing in the direction of increasing entropy ${\cal S}$
and the entropy is at a minimum value at $t=0$, solving the
conundrum of the Second Law of Thermodynamics. The Hubble radius
$H^{-1}(t)$ is infinite at $t=0$ removing the curvature and
particle horizons. The negative pressure $p_G$ generated by the
scalar field $\xi$ at $t\sim 0$ can produce quantum spontaneous
creation of particles explaining the origin of matter and
radiation.
\end{abstract}
\vskip 0.2 true in e-mail: john.moffat@utoronto.ca


\section{Introduction}

The singularity in standard cosmology at $t=0$ heralds the
breakdown of physics at the big bang. This has been a
long-standing problem in cosmology that prevents a rational
explanation for the onset of the beginning of the universe. We
know from the afterglow of the cosmic microwave background (CMB)
radiation with the uniform temperature $\sim 2.73$~K that,
according to the Friedmann-Lema\^{i}tre-Robertson-Walker (FLRW)
cosmology, there had to be a hot beginning to the universe. In the
following, we shall develop a non-singular cosmology using the
effective classical action and field equations based on modified
gravity (MOG)~\cite{Moffat,Moffat2,Moffat3}.

An initial value problem occurs at the beginning of the universe
at the big bang $t=0$, arising from one of the most basic laws of
physics: the second law of thermodynamics. This problem has been
addressed by Penrose~\cite{Penrose,Penrose2}, who has proposed a
solution based on making the Weyl curvature tensor vanish at
$t=0$. He also proposed that the universe went through a series of
conformally invariant cycles~\cite{Penrose3}.

We do not wish to contemplate a violation of the second law of
thermodynamics, which is as cherished as the law of conservation
of energy. The problem with the second law of thermodynamics
arises because the entropy ${\cal S}$ of the universe increases
as time increases from $t=0$, and accordingly the disorder or lack
of speciality increases as the universe expands. Therefore, the
initial state of the universe was the most special state of all.

In the following, we shall pursue a non-singular cosmological model
which can resolve initial value problems such as the horizon problem
and the entropy conundrum. We shall study the MOG cosmology using
the effective classical scalar-tensor-vector gravity
(STVG)~\cite{Moffat2,Moffat3}. This MOG with its variations of $G$,
the vector field $\phi_\mu$, the coupling strength $\omega$ and its
effective mass $\mu$ leads to a satisfactory description of galaxy
rotation curves, the mass profiles of X-ray clusters of
galaxies~\cite{Brownstein,Brownstein2}, the bullet cluster
1E0657-56~\cite{Brownstein3} and to the WMAP data for the CMB, the
matter power spectrum and the supernovae data~\cite{Toth} without
undetected dark matter and dark energy.

\section{MOG Field Equations and Action Principle}

The gravitational field equations are given by~\cite{Moffat2}:
\begin{equation}
\label{Einsteineqs}G_{\mu\nu}-g_{\mu\nu}\Lambda+Q_{\mu\nu}=8\pi
G_N\xi T_{\mu\nu},
\end{equation}
where $G_{\mu\nu}=R_{\mu\nu}-\frac{1}{2}g_{\mu\nu}R$, $\xi(x)$
denotes the scalar field describing the variation of the
gravitational ``constant'' and $G=G_N\xi(x)$ where $G_N$ denotes
Newton's constant. We have chosen units with $c=1$. We adopt the metric signature
$\eta_{\mu\nu}={\rm diag}(1,-1,-1,-1)$ where $\eta_{\mu\nu}$ is
the Minkowski spacetime metric, $R=g^{\mu\nu}R_{\mu\nu}$ and
$\Lambda$ denotes the cosmological constant.

We have
\begin{equation}
Q_{\mu\nu}=\xi\biggl[\nabla^\alpha\nabla_\alpha\biggl(\frac{1}{\xi}\biggr)
g_{\mu\nu}-\nabla_\mu\nabla_\nu\biggl(\frac{1}{\xi}\biggr)\biggr],
\end{equation}
where $\nabla_\mu$ denotes the covariant derivative with respect to
the metric $g_{\mu\nu}$. The quantity $Q_{\mu\nu}$ results from a
boundary contribution arising from the presence of second
derivatives of the metric tensor in $R$ in the action. These
boundary contributions are equivalent to those that occur in
Brans-Dicke gravity theory~\cite{Dicke}.

In the original formulation of MOG~\cite{Moffat2}, the vector
field $\phi_\mu$ satisfies massive Maxwell-Proca-type field
equations. We shall generalize these field equations to massive
Yang-Mills field equations of the form:
\begin{equation}
D_\nu B^{a\mu\nu}+\frac{\partial
V(\phi)}{\partial\phi_\mu^a}+\frac{1}{\omega}\Delta_\nu\omega
B^{a\mu\nu}=-\frac{1}{\omega}J^{a\mu}.
\end{equation}
Here, $B^{a\mu\nu}$ denotes the Yang-Mills field:
\begin{equation}
B^{a\mu\nu}=\partial_\mu\phi^a_\nu-\partial_\nu\phi^a_\mu +
C^{abc}\phi^b_\mu\phi^c_\nu,
\end{equation}
and we have included contributions from the variation of the
effective Yang-Mills coupling strength $\omega$. Moreover, $D_\mu$
denotes the Yang-Mills covariant derivative acting on a Dirac
field $\Psi(x)$:
\begin{equation}
D_\mu\Psi(x)=[\partial_\mu+\phi_\mu(x)]\Psi(x),
\end{equation}
where
\begin{equation}
\phi_\mu(x)=-i\omega\phi_\mu^a(x)T^a,\quad B^{\mu\nu}(x)=-i\omega
B^{a\mu\nu}(x)T^a.
\end{equation}
Here, $T^a$ are the matrix representations of the generators of
the $U(n)$ symmetry group.

In the absence of contributions from the potential $V(\phi)$ the
action for the Yang-Mills field $\phi_\mu$ is invariant under the
gauge transformations
\begin{equation}
\phi_\mu^{'}(x)=U(x)\phi_\mu(x)U^{-1}(x)-[\partial_\mu
U(x)]U^{-1}(x).
\end{equation}
However, in general there will be contributions in the potential
$V(\phi)$ from a mass term $-(1/2)\mu^2\phi^\mu\phi_\mu$ which
will break the gauge symmetry. This mass term can be obtained from
a spontaneous symmetry breaking of the action.

Our action takes the form
\begin{equation}
S=S_{\rm Grav}+S_\phi+S_S+S_M,
\end{equation}
where
\begin{equation}
S_{\rm Grav}=\frac{1}{16\pi}\int d^4x\sqrt{-g}\biggl[
\frac{1}{G}(R+2\Lambda)\biggr],
\end{equation}
\begin{equation}
S_\phi=-\int
d^4x\sqrt{-g}\biggl[\omega\biggl(\frac{1}{4}B^{\mu\nu}B_{\mu\nu} +
V(\phi)\biggr)\biggr],
\end{equation}
and
\begin{equation}
\label{Saction} S_S=\int
d^4x\sqrt{-g}\biggl[\frac{1}{G^3}\biggl(\frac{1}{2}g^{\mu\nu}\nabla_\mu
G\nabla_\nu G-V(G)\biggr)
$$ $$
+\frac{1}{G}\biggl(\frac{1}{2}g^{\mu\nu}\nabla_\mu\omega
\nabla_\nu\omega-V(\omega)\biggr)
+\frac{1}{\mu^2G}\biggl(\frac{1}{2}g^{\mu\nu}\nabla_\mu\mu\nabla_\nu\mu
-V(\mu)\biggr)\biggr].
\end{equation}
Moreover, $V(\phi)$ denotes a potential for the vector field
$\phi^\mu$, while $V(G), V(\omega)$ and $V(\mu)$ denote the three
potentials associated with the three scalar fields $G(x),\omega(x)$
and $\mu(x)$, respectively.

The current conservation law is of the form
\begin{equation}
D_\mu J^{a\mu}=0.
\end{equation}
The non-abelian charge associated with the fermion current density
is no longer a constant of the motion, for the gauge field
$\phi_\mu$ is no longer neutral with respect to the ``fifth
force'' charge and their is a non-linear coupling between the
$\phi_\mu$ fields, leading to a non-zero $\phi_\mu$ charge current
density. This means that the fermions carrying the non-abelian
fifth force charge obey a non-linear relation between the charges.
In the Maxwell-Proca version of MOG, which is invariant under
$U(1)$ abelian gauge transformations for massless $\phi_\mu$
fields, the fermion fifth force point charges add linearly. The
non-linear relation between fermion point charges in the
Yang-Mills description of the fifth force can have important
implications for the explanation of astrophysical
data~\cite{Brownstein,Brownstein2,Brownstein3}.

The total energy-momentum tensor is given by
\begin{equation}
T_{\mu\nu}=T_{M\mu\nu}+T_{\phi\mu\nu}+T_{S\mu\nu},
\end{equation}
where $T_{M\mu\nu}$, $T_{\phi\mu\nu}$ and $T_{S\mu\nu}$ denote the
energy-momentum tensor contributions of ordinary matter, the
$\phi_\mu$ field and the scalar fields $\xi$, $\omega$ and $\mu$,
respectively. We have
\begin{equation}
T_{\phi\mu\nu}=\omega\biggl[{B_\mu}^\alpha B_{\nu\alpha}
-g_{\mu\nu}\biggl(\frac{1}{4}B^{\rho\sigma}B_{\rho\sigma}
+V(\phi)\biggr)+2\frac{\partial V(\phi)}{\partial
g^{\mu\nu}}\biggr].
\end{equation}
The $\xi(x)$ field yields the energy-momentum tensor:
\begin{equation}
T_{\xi\mu\nu}=-\frac{1}{G_N\xi^3}\biggl[\nabla_\mu \xi\nabla_\nu
\xi-2\frac{\partial V(\xi)}{\partial g^{\mu\nu}}
-g_{\mu\nu}\biggl(\frac{1}{2}\nabla_\alpha\xi \nabla^\alpha
\xi-V(\xi)\biggr)\biggr],
\end{equation}
where $V(\xi)$ denotes a potential for the scalar field $\xi$.

From the Bianchi identities
\begin{equation}
\nabla_\nu G^{\mu\nu}=0,
\end{equation}
and from the field equations (\ref{Einsteineqs}), we obtain
\begin{equation}
\label{conservation} \nabla_\nu T^{\mu\nu}+\frac{1}{\xi}\nabla_\nu
\xi T^{\mu\nu}-\frac{1}{8\pi G_N\xi}\nabla_\nu Q^{\mu\nu}=0.
\end{equation}

The scalar field $\xi(x)$ satisfies the field equations
\begin{equation}
\label{Gequation} \nabla_\alpha\nabla^\alpha
\xi+V'(\xi)+N(\xi)=\frac{1}{2}G_N\xi^2\biggl(T+\frac{\Lambda}{4\pi
G_N\xi}\biggr),
\end{equation}
where
\begin{equation}
N(\xi)=-\frac{3}{\xi}\biggl(\nabla_\alpha \xi\nabla^\alpha \xi
+V(\xi)\biggr)+\frac{1}{16\pi}\xi^2\nabla_\alpha\nabla^\alpha\biggl(\frac{1}{\xi}\biggr)
$$ $$
+\xi\biggl(\frac{1}{2}\nabla_\alpha\omega\nabla^\alpha\omega-V(\omega)\biggr)
+\frac{\xi}{\mu^2}\biggr(\frac{1}{2}\nabla_\alpha\mu\nabla^\alpha\mu-V(\mu)\biggr),
\end{equation} and $T=g^{\mu\nu}T_{\mu\nu}$. Similar field
equations hold for the scalar fields $\omega(x)$ and
$\mu(x)$~\cite{Moffat2}.

We observe that our field equations (\ref{Gequation}) for the
variation of $G$ contain a potential $V(\xi)$, which is absent in
standard Brans-Dicke gravity~\cite{Dicke}. Moreover, the
conservation of energy equation (\ref{conservation}) is more
general than in Brans-Dicke gravity in which $\nabla_\nu
T^{\mu\nu}_M=0$ is imposed from the outset.

The trace free Weyl curvature tensor is defined by
\begin{equation}
\label{Weyl}
C_{\mu\nu\rho\sigma}=R_{\mu\nu\rho\sigma}-g_{\mu[\rho}
R_{\sigma]\nu} -g_{\nu[\rho}
R_{\sigma]\mu}+\frac{1}{3}Rg_{\mu[\sigma} g_{\rho]\nu}.
\end{equation}
Under a conformal transformation of the metric:
\begin{equation}
{\tilde g}^{\mu\nu}=\Omega^{-2} g^{\mu\nu},
\end{equation}
the Weyl tensor is unchanged:
\begin{equation}
{\tilde C}_{\mu\nu\rho}^\sigma=C_{\mu\nu\rho}^\sigma.
\end{equation}
The field equations (\ref{Einsteineqs}) are not conformally
invariant due to the non-vanishing trace of the energy-momentum
tensor, $T\not= 0$.

\section{Non-Singular Cosmology}

Let us now consider a cosmological solution to our MOG theory. We
adopt a homogeneous and isotropic FLRW background geometry with
the line element
\begin{equation}
\label{metric}
ds^2=dt^2-a^2(t)\biggl(\frac{dr^2}{1-kr^2}+r^2d\Omega^2\biggr),
\end{equation}
where $d\Omega^2=d\theta^2+\sin^2\theta d\phi^2$ and $k=0,-1,+1$ for
a spatially flat, open and closed universe, respectively. Due to the
symmetry of the FLRW background spacetime, we have $\phi_0\not= 0$,
$\phi_i=0$ and $B_{\mu\nu}=0$. The metric (\ref{metric}) is
conformally flat to a Minkowski spacetime metric.

We define the energy-momentum tensor for a perfect fluid by
\begin{equation}
T^{\mu\nu}=(\rho+p)u^\mu u^\nu-pg^ {\mu\nu},
\end{equation}
where $u^\mu=dx^\mu/ds$ is the 4-velocity of a fluid element and
$g_{\mu\nu}u^\mu u^\nu=1$. Moreover, we have
\begin{equation}
\rho=\rho_M+\rho_\phi+\rho_S,\quad p=p_M+p_\phi+p_S,
\end{equation}
where $\rho_i$ and $p_i$ denote the components of density and
pressure associated with the matter, the $\phi^\mu$ field and the
scalar fields $\xi$, $\omega$ and $\mu$, respectively.

The modified Friedmann equations take the
form~\cite{Moffat,Moffat2,Moffat3}:
\begin{equation}
\label{Friedmann1} \frac{\dot
a^2(t)}{a^2(t)}+\frac{k}{a^2(t)}=\frac{8\pi
G_N\xi(t)\rho(t)}{3}+f(t)+\frac{\Lambda}{3},
\end{equation}
\begin{equation}
\label{Friedmann2} \frac{{\ddot a}(t)}{a(t)}=-\frac{4\pi
G_N\xi(t)}{3}[\rho(t)+3p(t)]+h(t) +\frac{\Lambda}{3},
\end{equation}
where $\dot a=da/dt$ and
\begin{equation}
\label{fequation} f(t)=\frac{\dot a(t)}{a(t)}\frac{{\dot
\xi(t)}}{\xi(t)},
\end{equation}
\begin{equation}
\label{hequation} h(t)=\frac{1}{2}\biggl(\frac{{\ddot
\xi(t)}}{\xi(t)}-2\frac{{{\dot \xi}^2(t)}}{\xi^2(t)}+\frac{\dot
a(t)}{a(t)}\frac{{\dot \xi(t)}}{\xi(t)}\biggr).
\end{equation}
The conservation law for matter is given by
\begin{equation}
\label{conserveeq} \dot\rho+3\frac{d\ln
a}{dt}(\rho+p)+\rho\frac{\dot\xi}{\xi}+{\cal I}=0,
\end{equation}
where
\begin{equation}
\label{Iequation} {\cal I}=\frac{3}{8\pi G_N\xi a}(2{\dot a}f+a\dot
f-2{\dot a}h).
\end{equation}

Let us make the simplifying approximation for the equations
(\ref{Gequation}) with $\Lambda=0$:
\begin{equation}
\label{approxGeq} {\ddot{\xi}}+3H{\dot{\xi}}+V'(\xi)
=\frac{1}{2}G_N\xi^2(\rho-3p).
\end{equation}
An approximate solution for $\xi$ in terms of a given potential
$V(\xi)$ and for given values of $\rho$ and $p$ can be obtained from
(\ref{approxGeq}).

We will assume the following conditions for a non-singular
solution to occur at $t=0$:
\begin{enumerate}

\item $ a(0)>0$\, ({\rm hence, no singularity at}\, $t=0$),

\item $ {\dot a}(0)=0$\, ({\rm bounce or static solution}),

\item $ \xi(0) > 0$,

\item $ {\ddot a}(0) > 0$\, ({\rm bounce solution}),

\item $\rho(0)=0$\, {\rm and}\, $\dot\rho(0)=0$\, ({\rm from which
follows}\, $\rho(t) > 0$\, {\rm for}\, $t > 0$\, {\rm and}\, $t
<0$.)

\end{enumerate}

By setting the cosmological constant and the curvature constant to
zero, $\Lambda=k=0$, we get from the generalized Friedmann equations
(\ref{Friedmann1}) and (\ref{Friedmann2}) for a singularity-free
universe:
\begin{equation}
\xi(0)\rho(0)=0.
\end{equation}
Because $\xi(0)\neq 0$, otherwise the Friedmann equations become
singular, we must have $\rho(0)=0$.

We require that ${\dot\rho}(0)=0$ to avoid $\rho$ changing sign at
$t=0$. This means that the strong energy condition, $\rho \geq 0$,
is satisfied for all $t$. Moreover, the conservation law
(\ref{conserveeq}) leads to ${\cal I}(0)=0$. From
(\ref{Iequation}) it follows that $a(0){\dot f}(0)=0$ and at
$t=0$:
\begin{equation}
\frac{d}{dt}\biggl(\frac{{\dot a}}{a}\frac{\dot\xi}{\xi}\biggr)
=\biggl(\frac{{\ddot a}}{a}-\frac{{\dot
a}^2}{a^2}\biggr)\frac{\dot\xi}{\xi}+\frac{{\dot
a}}{a}\biggl(\frac{\ddot\xi}{\xi}-\frac{{\dot\xi}^2}{\xi}\biggr)=0.
\end{equation}
Because ${\dot a}(0)=0$ and ${\ddot a}(0) > 0$, we have
$\dot\xi(0)=0$.

Let us now return to the second Friedmann equation
(\ref{Friedmann2}). We have from (\ref{hequation}) at $t=0$:
\begin{equation}
h=-\frac{1}{2}\xi{\ddot\Theta},
\end{equation}
where $\Theta(t)=1/\xi(t)$. Therefore, from the second Friedmann
equation we obtain at $t=0:$
\begin{equation}
\frac{{\ddot a}}{a}=-4\pi
G_N\xi\biggl(p+\frac{1}{2}\ddot\Theta\biggr).
\end{equation}
Let us write this equation in the form at $t=0$:
\begin{equation}
\label{Fried1} \frac{{\ddot a}}{a}=-4\pi G_N\xi (p_m+p_G),
\end{equation}
where $p_m$ denotes the matter pressure and we have for $\xi(0) >
0$ at $t=0$:
\begin{equation}
p_G=\frac{\ddot\Theta}{8\pi G_N}.
\end{equation}
We require that $\ddot\Theta < 0$ and $-p_G > p_m$ which gives
${\ddot a}(0) > 0$.

Spacetime at $t\sim 0$ is described by the conformally flat
Minkowski metric:
\begin{equation}
ds^2=dt^2-a^2(0)(dx^2+dy^2+dz^2).
\end{equation}
The conformal transformation
\begin{equation}
{\tilde g}_{\mu\nu}=\Omega^2 g_{\mu\nu},
\end{equation}
is well-defined for $\Omega =\sqrt{\xi}$, for $\xi$ is finite at
$t=0$. Quantum fluctuations of the $\xi$ field in the neighborhood
of $t=0$ cause the Minkowski spacetime to become unstable for
$t\sim 0$ generating an expansion of the universe as $t$ increases
in time.

From the first Friedmann equation (\ref{Friedmann1}) with
$k=\Lambda=0$, we obtain
\begin{equation}
\frac{\dot{a}(t)}{a(t)}=\biggl[\frac{8\pi
G_N\xi(t)\rho(t)}{3}+f(t)\biggr]^{1/2}.
\end{equation}
This equation has the formal solution
\begin{equation}
\label{asolution} a(t)=\exp\biggl\{\int_0^t dt'\biggl[\frac{8\pi
G_N\xi(t')\rho(t')}{3}+f(t')\biggr]^{1/2}+A\bigg\},
\end{equation}
where $A$ is a constant of integration and we have
$a(0)=\exp(A)={\rm constant}$. We require that a closed solution
of (\ref{asolution}) becomes the radiation dominant solution
$a(t)\propto t^{1/2}$ as $t$ increases from $t=0$.

The Hawking-Penrose theorems~\cite{Penrose4,Hawking,Ellis} prove
that a singularity must occur at $t=0$ in classical GR when the weak
and strong energy conditions are satisfied. The weak and strong
energy conditions are satisfied when
\begin{equation}
\label{weakEC} T_{\mu\nu}U^\mu U^\nu \geq 0,
\end{equation}
and
\begin{equation}
\label{strongEC} (T_{\mu\nu}-\frac{1}{2}g_{\mu\nu}T)U^\mu U^\nu
\geq 0,
\end{equation}
where $U^\mu$ is a timelike or null vector. The conditions
(\ref{weakEC}) and (\ref{strongEC}) hold when $\rho \geq 0$ and
$\rho+3p \geq 0$, respectively. We have for $p_G < -p_m$ in the
second Friedmann equation (\ref{Fried1}) that $\rho+3p$ is
negative for $t\sim 0$ violating the strong energy condition. We
conclude from this that our non-singular cosmological solution is
consistent with the Hawking-Penrose theorems, for the violation of
the strong energy condition invalidates the no-go singularity
theorems.

\section{Evolution of Non-Singular Cosmology}

Our cosmological solution derived from our MOG satisfies a
non-singular solution in which the universe has $\rho(0)=0$ and
${\dot a}(0)={\dot\xi}(0)=0$ and it satisfies the radiation
dominated solution of the FLRW type for $t
> 0$. The negative pressure caused by $p_G < -p_m$ at $t\sim 0$ due to the scalar
field $\xi$ can lead to a quantum mechanical production of
particles. Such a scenario has been investigated by Parker and
Fulling~\cite{Parker} and Brout, Englert and
Gunzik~\cite{Englert}.

The Weyl curvature tensor defined in (\ref{Weyl}) vanishes
identically in the homogeneous and isotropic
 FLRW universe, $C_{\mu\nu\rho\sigma}=0$. Therefore, the gravitational degrees
 of freedom vanish and due to the vanishing of the matter density, $\rho(0)=0$, the
entropy of the universe ${\cal S}$ is at a minimum in the
neighborhood of $t=0$. As $\rho(t)$ grows from zero as $|t|$
increases for either $t\rightarrow -\infty$ or $t\rightarrow
+\infty$, the entropy will increase and obey the second law of
thermodynamics as the universe expands. However, this scenario
requires that {\it the thermodynamical arrow of time reverses} at $t=0$ and always
points in the direction of increasing entropy. Thus, an observer
will see a universe with {\it increasing} entropy as $t\rightarrow
-\infty$. The solution for $a(t)$ given by (\ref{asolution}) can
be time-symmetric or time-asymmetric depending on the particular
closed-form solution obtained from $\rho(t)$ and $\xi(t)$. At
$t\sim 0$ the gravitational field and the negative pressure $p_G <
0$ can produce matter as pairs of particles are spontaneously
created from vacuum fluctuations.

For our non-singular universe the Hubble parameter $H(t)={\dot
a}(t)/a(t)$ obeys $H(0)=0$ and the universe stops expanding at
$t=0$. The epoch at $t \geq 0$ when the universe begins to expand
consists of a hot radiation plasma and due to the red shift of the
radiation predicts the observed uniform temperature of the CMB.

The proper particle horizon size is given by
\begin{equation}
\label{horizon} d_H(t)=a(t)\int_{-\infty}^{t}\frac{dt'}{a(t')}
=a(t)\int_{-\infty}^{t}\frac{da(t')}{a^2(t')H(t')}.
\end{equation}
The horizon size $d_H(t)$ becomes infinite as the universe evolves
from $t=-\infty$ through $t=0$ to $t\rightarrow\infty$. This
allows photons to be thermalized through repeated particle
collisions and there is no problem with causality, predicting the
uniform CMB temperature.

\section{Conclusions}

We have obtained from MOG a non-singular description of the large
scale structure of spacetime. As the universe evolves from
$t=-\infty$ there are no cosmological singularities and there is
no ``big bang'' at the putative origin of the universe at $t=0$.
The combined physics of quantum production of matter and MOG
explain the origin of matter in the universe shortly after $t=0$
and the hot radiation plasma and the subsequent expansion of the
universe has a similar evolution as in the big bang model. We have
seen that it is possible in the singularity-free MOG cosmology to
solve the conundrum of the entropy problem, for the gravitational
degrees of freedom are zero due to the vanishing of the Weyl
curvature, and, because $\rho(0)=0$ the entropy ${\cal S}$ is at a
minimum for $t\sim 0$. The arrow of time for negative and positive
$t$ must follow the increase in entropy from the minimum at $t=0$.
This is guaranteed by a reversal of the arrow of time at $t=0$,
which avoids a violation of the second law of thermodynamics.
Matter is spontaneously created at $t\sim 0$ due to the
gravitational field and quantum fluctuations of the vacuum with
negative $p_G$, explaining the origin of matter and radiation.
\vskip 0.2 true in {\bf Acknowledgments} \vskip 0.2 true in

This work was supported by the Natural Sciences and Engineering
Research Council of Canada. I thank Martin Green and  Viktor Toth
for helpful and stimulating discussions.

\end{document}